\documentclass[a4paper, amsfonts, amssymb, amsmath, reprint,aps, prl,superscriptaddress,showkeys, nofootinbib,twoside,twocolumn]{revtex4-2}

\usepackage{parskip}
\usepackage[english]{babel}
\usepackage[utf8]{inputenc}
\usepackage{bbold,dsfont}
\usepackage[mathscr]{eucal}
\usepackage{braket}
\usepackage{comment}
\usepackage{subfigure}
\usepackage[colorinlistoftodos, color=green!40, prependcaption]{todonotes}
\usepackage{lineno}
\usepackage{nameref}
\usepackage{varioref}
\usepackage{hyperref}
\usepackage{cleveref}
\graphicspath{{./pics/}, [interpolate,keepaspectratio]}
\usepackage[ruled,vlined]{algorithm2e}

\begin{document}

\title{
Autonomous Floquet Engineering of Bosonic Codes via Reinforcement Learning}

\author{Zheping Wu}
\affiliation{%
School of Computer Science, Northwestern Polytechnical University, Xi’an 710129, China 
}

\author{Lingzhen Guo}
\affiliation{Center for Joint Quantum Studies and Department of Physics, School of Science, Tianjin University,
Tianjin 300072,  China}
\affiliation{Tianjin Key Laboratory of Low Dimensional Materials Physics and Preparing Technology, Department
of Physics, Tianjin University, Tianjin 300354, China}
\author{Haobin Shi}
\email{shihaobin@nwpu.edu.cn}
\affiliation{%
School of Computer Science, Northwestern Polytechnical University, Xi’an 710129, China 
}

\author{Wei-Wei Zhang}
\email{Corresponding author: wei-wei.zhang@nwpu.edu.cn}

\affiliation{%
School of Computer Science, Northwestern Polytechnical University, Xi’an 710129, China 
}
\date{\today} 

\begin{abstract}
Bosonic codes represent a promising route toward quantum error correction in continuous-variable systems, with direct relevance to experimental platforms such as circuit QED and optomechanics. However, their preparation and stabilization remain highly challenging, requiring ultra-precise control of nonlinear interactions to create entangled superpositions, suppress decoherence, and mitigate dynamic errors. Here, we introduce a reinforcement-learning-assisted Floquet engineering approach for the autonomous preparation of bosonic codes that is general, efficient, and noise-resilient.  By leveraging machine learning to optimize 
Floquet driving parameters, our method achieves over two orders of magnitude reduction in evolution time—requiring only about one percent of that in conventional adiabatic schemes—while maintaining high-fidelity state generation even under strong dissipative and dephasing noise. This approach not only demonstrates the power of artificial intelligence in quantum control, but also establishes a scalable and experimentally feasible route toward fault-tolerant bosonic quantum computation. Beyond the specific application to bosonic code preparation, our results suggest a general paradigm for integrating machine learning and Floquet engineering to overcome decoherence challenges in next-generation quantum technologies.
\end{abstract}

\maketitle
\section{Introduction}
Bosonic codes are a class of quantum error correction (QEC) codes  that encode quantum information into the quantum states of continuous-variable (CV) systems, e.g., cavities and harmonic oscillators~\cite{Chuang-PhysRevA.56.1114,Cochrane-PhysRevA.59.2631,Gottesman-PhysRevA.64.012310}. Unlike traditional multi-qubits based QEC codes such as surface codes~\cite{KITAEV20032,PhysRevA.86.032324}, bosonic codes exploit the infinite-dimensional Hilbert space of a single bosonic system and encode logical qubits into the non-classical states of electromagnetic fields in microwave cavities or optical modes. By embedding quantum information into proper non-Gaussian codes states, e.g., cat codes~\cite{Cochrane-PhysRevA.59.2631,dodonov1974even,mirrahimi2014dynamically,bergmann2016quantum},  Gottesman-Kitaev-Preskill (GKP) codes~\cite{Gottesman-PhysRevA.64.012310}, binomial codes~\cite{michael2016new} or vertex codes~\cite{Hu2025}, bosonic codes can be tailored to correct dominant error sources in the experiments, such as photon loss and phase fluctuations. For instance, cat codes intrinsically protect against photon losses by leveraging parity symmetry~\cite{Jayashankar-PhysRevResearch.4.023034}, while GKP codes correct diffusion of quadratures in phase space through periodic stabilization~\cite{Arne-PRXQuantum.2.020101}. Due to the reduced physical overhead~\cite{Sabo-PRXQuantum.5.040302}, bosonic codes provide a hardware efficient scalable solution for fault-tolerant quantum computation in various platforms such as  superconducting circuits, trapped ions and optical systems.

\begin{figure*}
    \centering
    \includegraphics[width=1.4\columnwidth]{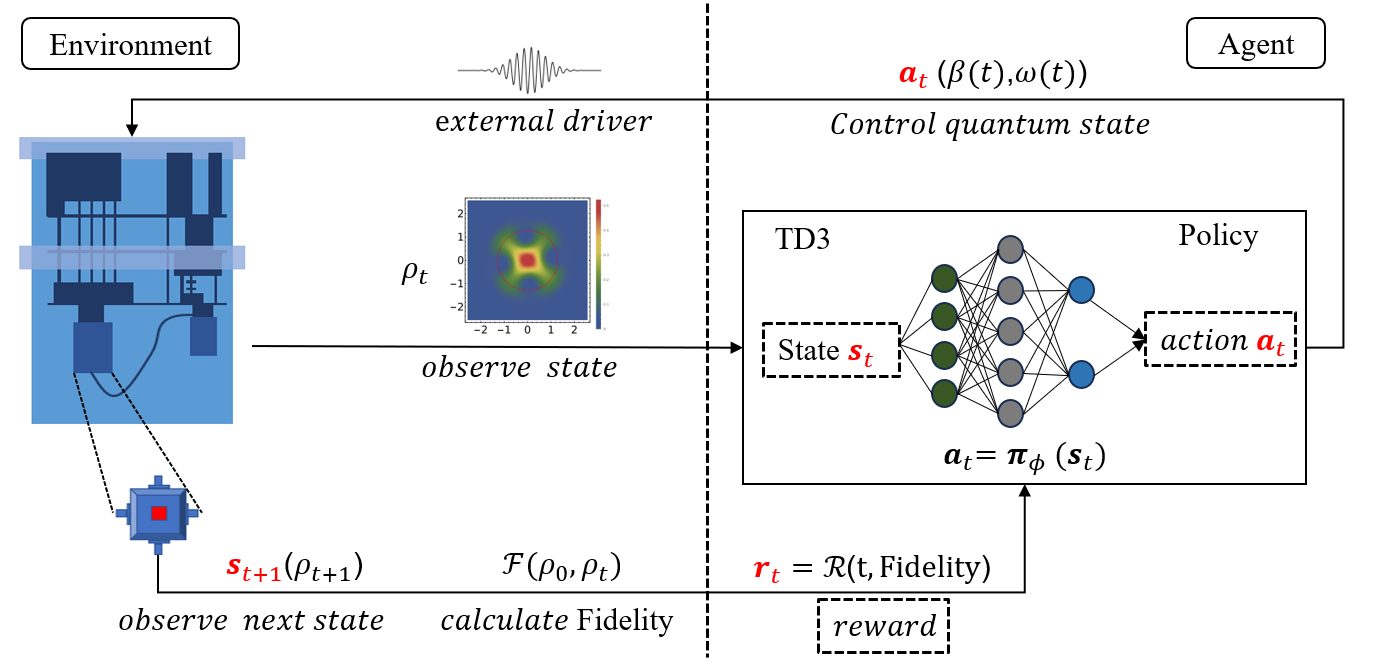}
         \caption{A sketch of our reinforcement learning process for the preparation of the bosonic cat states,  where the external driving field is dynamically updated by the agent based on the instantaneous fidelity of the evolution state.}
         \label{fig:sketch}
\end{figure*}

A universal quantum computing scheme applicable to the cat and binomial codes has been proposed, including the code-agnostic entangling gate that can be used to interface different rotational symmetric encodings~\cite{Arne-PhysRevX.10.011058}. The experimental realizations of quantum computation with bosonic codes has been reported in superconducting circuits~\cite{michael2016new,hu2019quantum,ma2021quantum,cai2021bosonic,brady2024advances}, neutral atoms~\cite{Bohnmann-PhysRevA.111.022432}. 

Despite the promising advantages from the theory, bosonic codes face realistic experimental challenges. Both the encoding and error correcting processes continuously consume bosonic code states, it remains technically demanding to prepare high-fidelity non-classical states (e.g., large cat states or GKP grids) efficiently.
The preparation of bosonic code states can be achieved via universal gate operations like the cubic phase gate~\cite{PhysRevResearch.6.023332}  and the selective number-dependent arbitrary phase (SNAP) gate~\cite{PhysRevA.92.040303,guo2024engineering}.
Alternatively, passive control strategies were proposed to engineer code states  based on Hamiltonian engineering and adiabatic ramp~\cite{Arne-PhysRevLett.132.130605,guo-PhysRevLett.132.023602}. However,  the methods based on the adiabatic approach are slow (thousands of driving periods) and noise sensitive, thus require a high requirement for system coherence.

The rapid advancement of artificial intelligence (AI) technologies has greatly accelerated progress across numerous scientific domains, including quantum technology. Among the key challenges in this field is the efficient and accurate preparation of quantum states. Recent studies have proposed various reinforcement learning (RL) approaches to address this problem, demonstrating notable advantages such as improved control efficiency, robustness, and adaptability. 
For instance, AI-assisted methods have been employed for the preparation of Dicke states~\cite{Guo-PhysRevLett.126.060401}, many-body phase transition states~\cite{Bukov-PhysRevX.8.031086}, quantum thermal and prethermal quantum states~\cite{Baba-PhysRevApplied.19.014068}, Fock state preparation with the weak nonlinear measurements~\cite{porotti2022deep}, and optimized GKP codes generation~\cite{Zeng-PhysRevLett.134.060601}. In this work, we present an efficient Floquet-engineering scheme for the preparation of bosonic code states using reinforcement learning. Our approach reduces the total preparation time for a target code state from the previously required thousands of driving periods~\cite{Arne-PhysRevLett.132.130605,guo-PhysRevLett.132.023602} to only tens of periods, while maintaining strong robustness against photon loss and dephasing noise.

\section{Model}
\paragraph{\textbf{Preparation of multi-components cat codes with Floquet engineering}}
Multi-components cat codes belong to the class of rotational bosonic codes that exhibit discrete rotational symmetries in the phase space of a bosonic mode~\cite{Arne-PhysRevX.10.011058}. 
The rotational symmetry enforces a spacing of the
code words in Fock space, which underpins the robustness of the codes to photon loss errors.  
The logical code words for any $q$-fold rotational  symmetry code can be constructed from  superposition of a discretely-rotated normalized primitive state $|\Theta\rangle$. The two single-quibt 
states $\ket{0_{q,\Theta}}$ and $|1_{q,\Theta}\rangle$ satisfy the relationship and $R_q|j_{q,\Theta}\rangle=(-1)^{j}|j_{q,\Theta}\rangle$
are given by
\begin{align}
    \ket{0_{q,\Theta}}&:=\frac{1}{\mathcal{N}_0} \sum_{m=0}^{2q-1} e^{\text{i}(m\pi/q)\hat{n}}\ket{\Theta}
    \label{eq:q-flod-cat0}\\
    \ket{1_{q,\Theta}}&:=\frac{1}{\mathcal{N}_1} \sum_{m=0}^{2q-1} (-1)^m e^{\text{i}(m\pi/q)\hat{n}}\ket{\Theta}.
    \label{eq:q-flod-cat1}
\end{align}
Here, $\mathcal{N}_i$ are normalization constants, $\hat{n}$ is photon number operator and $R_q=e^{\text{i}(\pi/q)\hat{n}}$ is the $q$-fold rotational operator. In this work, we set $q=4$ as an example to engineer four-legged cat states that are able to correct single photon loss error~\cite{gertler2021protecting,guo2024engineering,Guo-PhysRevLett.126.060401,xu2024perturbative}.

The preparation of the bosonic codes based on Floquet engineering, i.e.,  the system is driven by an external field periodically, has been developed independently by Guo et al ~\cite{guo-PhysRevLett.132.023602,xu2024perturbative,guo2024engineering} and Grimsmo et al~\cite{Arne-PhysRevLett.132.130605}. Both of the two schemes are based on the adiabatic ramping process that is guaranteed by setting the driving strength and frequency to follow an empirical sigmoid function.  
In Ref.~\cite{Arne-PhysRevLett.132.130605}, the authors utilized the kicked harmonic oscillator method, with the truncated Fourier series decomposition of the delta-function driving at a finite number of harmonics, to prepare GKP state.  While in the Refs.~\cite{guo-PhysRevLett.132.023602,xu2024perturbative,guo2024engineering}, the authors developed a more general driving scheme based on the non-commutative Fourier transformation (NcFT) that can synthesize arbitrary target Hamiltonians and prepare arbitrary bosonic code states. Without loss of generality, we choose the latter to demonstrate our scheme. \\ 


\paragraph{\textbf{Reinforcement learning for the preparation of bosonic states}} 
We propose an efficient reinforcement-learning framework that enables the rapid and high-fidelity preparation of multi-component cat states. 
As shown in Fig.~\ref{fig:sketch}, we sketch the preparation process of the cat state that is realized in the superconducting cavity with circuits chip confined in a Dilution refrigerator. The system control is fulfilled via the classical CPU for the realization of the external driving field. 
Our reinforcement learning algorithm is implemented by the named agent as shown in Fig.~\ref{fig:sketch}. Specifically, the agent realizes the popular \textit{twin-delayed deep deterministic} (TD3) policy gradient algorithm~\cite{fujimoto2018addressing}, which has demonstrated stable learning process with high sample efficiency. TD3 is an off-policy actor-critic method for environments with a continuous action space.  Briefly, the TD3 algorithm achieves reliable performance by mitigating action-value overestimation through a double Q-learning structure. The robustness of the Q-function is further improved by smoothing the noise in the target policy, while delayed policy updates promote stable and convergent value estimation.  These features make TD3 an excellent candidate for quantum control tasks.

In this work, we integrate Floquet engineering with reinforcement learning to achieve efficient quantum state preparation, inspired by the concept of engineering a target Hamiltonian whose eigenstate corresponds to the desired quantum state in Ref.~\cite{guo-PhysRevLett.132.023602}. 
In Floquet theory, we can realize a given target Hamiltonian by applying external driving field to the superconducting cavity with the system Hamiltonian given by
\begin{equation}
    \hat{\mathcal{H}}(t)=\frac{\omega_0}{2}(\hat{p^2}+\hat{x}^2)+\beta(t)V_\gamma(x,\omega(t)t).
    \label{eq:RL-hamiltonian}
\end{equation} 
The external driving field $V_\gamma(t)$ can be  formally 
written as a superposition of sinusoidal potentials~\cite{guo-PhysRevLett.132.023602} as following, 
\begin{equation}
V_\gamma(x,t)=\int_0^{+\infty}A(k,\omega_0t)\cos[kx+\phi(k,\omega_0t)]dk,
    \label{eq:driving-potential}
\end{equation}
with time-varying amplitudes $A(k,\omega_0t)$ and frequencies $\varphi(k,\omega_0t)$ 
determined from the NcFT coefficients of target Hamiltonian,  i.e.,
$$A= k|f_T (k\cos{\tau}, k\sin{\tau})|,\ \  \varphi= \text{Arg} [f_T (k\cos{\tau}, k\sin{\tau} )].$$
The detailed analytical expression of the NcFT coefficient $f_T (k\cos{\tau}, k\sin{\tau})$ can be found in Ref.~\cite{guo-PhysRevLett.132.023602}.  

Without loss of generality, we prepare the $q$-fold cat states, cf. Eq.~(\ref{eq:q-flod-cat0}), that are the eigenstates of the following target Hamiltonian~\cite{guo-PhysRevLett.132.023602} 
\begin{equation}
\hat{H}_{F\gamma}^{(T)}=\frac{\beta}{|\alpha_0|^{2q}}e^{-\gamma \hat{a}^\dagger \hat{a}}(\hat{a}^{\dagger q}-\alpha_0^{*q})(\hat{a}^q-\alpha_0^q)e^{-\gamma \hat{a}^\dagger \hat{a}}.
\label{eq:target-hamiltonian}
\end{equation}
Here, $\alpha_0=1/\sqrt{2\lambda}$ is the coherent state number, and the factor $e^{-\gamma\hat{a}^{\dagger}\hat{a}}$
 with $\gamma>0$ is introduced to suppress the divergence of Hamiltonian in phase space for NcFT .

As shown in Fig.~\ref{fig:sketch}, the superconducting cavity evolves from the initial vacuum state under the driven Hamiltonian given by Eq.~(\ref{eq:RL-hamiltonian}).
The quality of preparation is evaluated with the fidelity $\mathcal{F}(\rho_0,\rho_\text{pre}(t))$ of the prepared state $\rho_\text{pre}(t)$ with respect to the target code state $\rho_0=\ket{\psi_0}\bra{\psi_0}$ with $\ket{\psi_0}=\ket{0_{q,\Theta}}$ given by Eq.~(\ref{eq:q-flod-cat0}) in our reinforcement learning scheme.  Within our reinforcement-learning framework for quantum state preparation, the driving amplitude $\beta(t)$ and frequency  $\omega(t)$ in Eq.~(\ref{eq:RL-hamiltonian}) act as variational parameters that determine the system’s evolution.  Our TD3 algorithm updates the parameters $\beta(t)$ and $\omega(t)$ according to the current value of the fidelity $\mathcal{F}(t)$, see more details in Section of Methods. In the following,  we list all the definitions used in our scheme.
\begin{itemize}

    \item {\bf Environment:} the Floquet dynamics of the  quantum system under the external driving field. 
     \item {\bf State $s_t$:} the density matrix $\rho(t)$ of the quantum system at evolution time $t$. 
       
   \item {\bf Action $a_t$:} the update in amplitude and frequency of the driving pulses causing the environment to transition from current state $s_t$ to the next state $s_{t+1}$.      

    \item {\bf Reward $ \mathcal{R}$:} the criterion for evaluating the quality of the action $a_t$ taken by policy ${\pi_\phi}$, which is calculated based on the fidelity between the current quantum state and the target state at time $t$. 
    %
    
    \item {\bf Agent:} the agent decides the action $a_t$ to be taken based on the  current observed density matrix,   converts the action into a parameter update strategy, and then applies it to the system environment for the continued quantum evolution. 
\end{itemize}

\begin{figure}[!h]
    \centering
    \includegraphics[width=0.9\columnwidth]{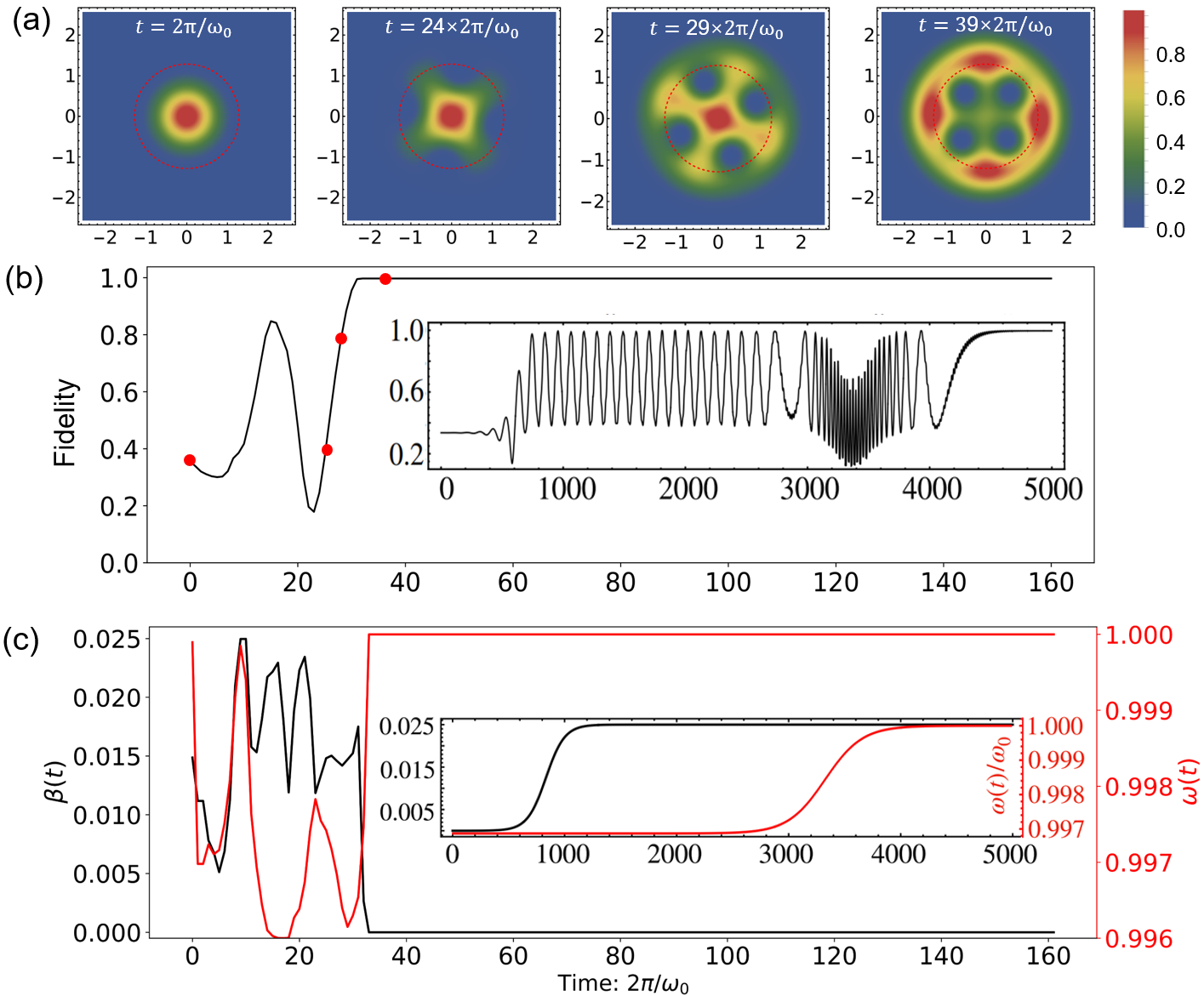}
               \caption{Preparation of 4-fold rotational bosonic cat state with reinforcement learning. (a) Snapshots of Husmi  Q-functions of prepared states $\bra{\alpha}\rho_t\ket{\alpha}$ at four different
time moments.  (b) Stroboscopic 
time evolution of the prepared state fidelity
$\mathcal{F}[\rho_0, \rho(t)]$, cf. Eq.~(\ref{eq:fidelity}), with respect to the 4-fold rotational target  bosonic state  $\rho_0=\ket{0_{q,\Theta}}\bra{0_{q,\Theta}}$ given by Eq.~(\ref{eq:q-flod-cat0}). The red dots indicate the corresponding time moments of the snapshots in (a). Inset: reproduced stroboscopic 
time evolution of fidelity of the prepared state using the adiabatic ramp method in  Ref.~\cite{guo-PhysRevLett.132.023602}.
(c) The machine learned driving amplitude $\beta(t)$ (black
curve) and the driving frequency $\omega(t)$ (red curve) with the preparation time $160×\times 2\pi/\omega_0$. Inset: reproduced ramp for the driving amplitude $\beta(t)$ (black
curve) and the driving frequency $\omega(t)$ (red curve) with the preparation time $5000×\times 2\pi/\omega_0$ in  Ref.~\cite{guo-PhysRevLett.132.023602}. \label{fig:evolution-process}         }
\end{figure}

\begin{figure}
    \centering
\includegraphics[width=0.97\columnwidth]{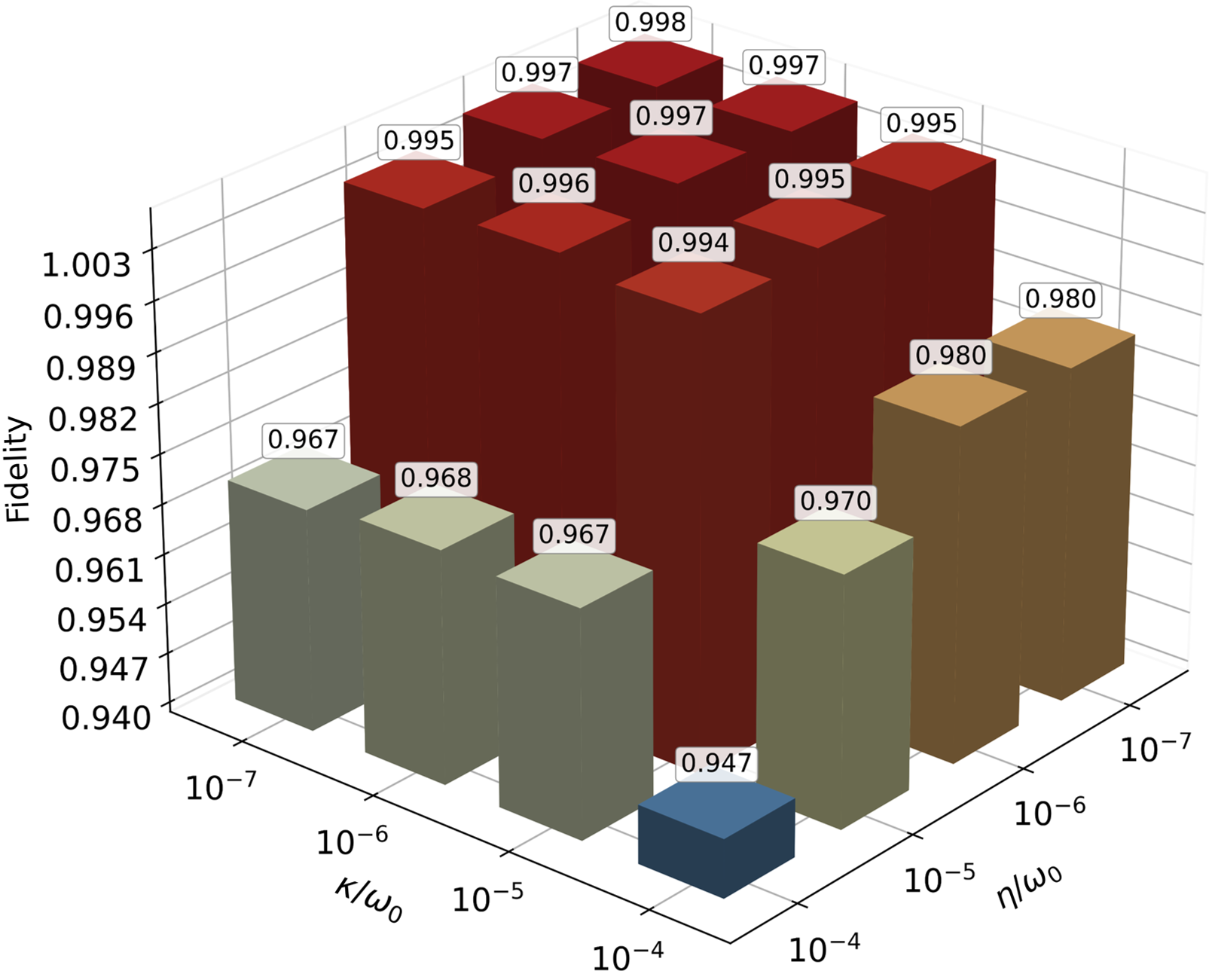}
         \caption{Fidelity of the prepared state with various levels of photon loss rate $\kappa$ and dephasing rate $\eta$ using the network trained with the lowest noise model, i.e. the case with ${\kappa = 10^{-7}\omega_0}$, ${\eta=10^{-7}\omega_0}$.}
         \label{fig:noise-test}
\end{figure}

\begin{figure}
    \centering
\includegraphics[width=1.0\columnwidth]{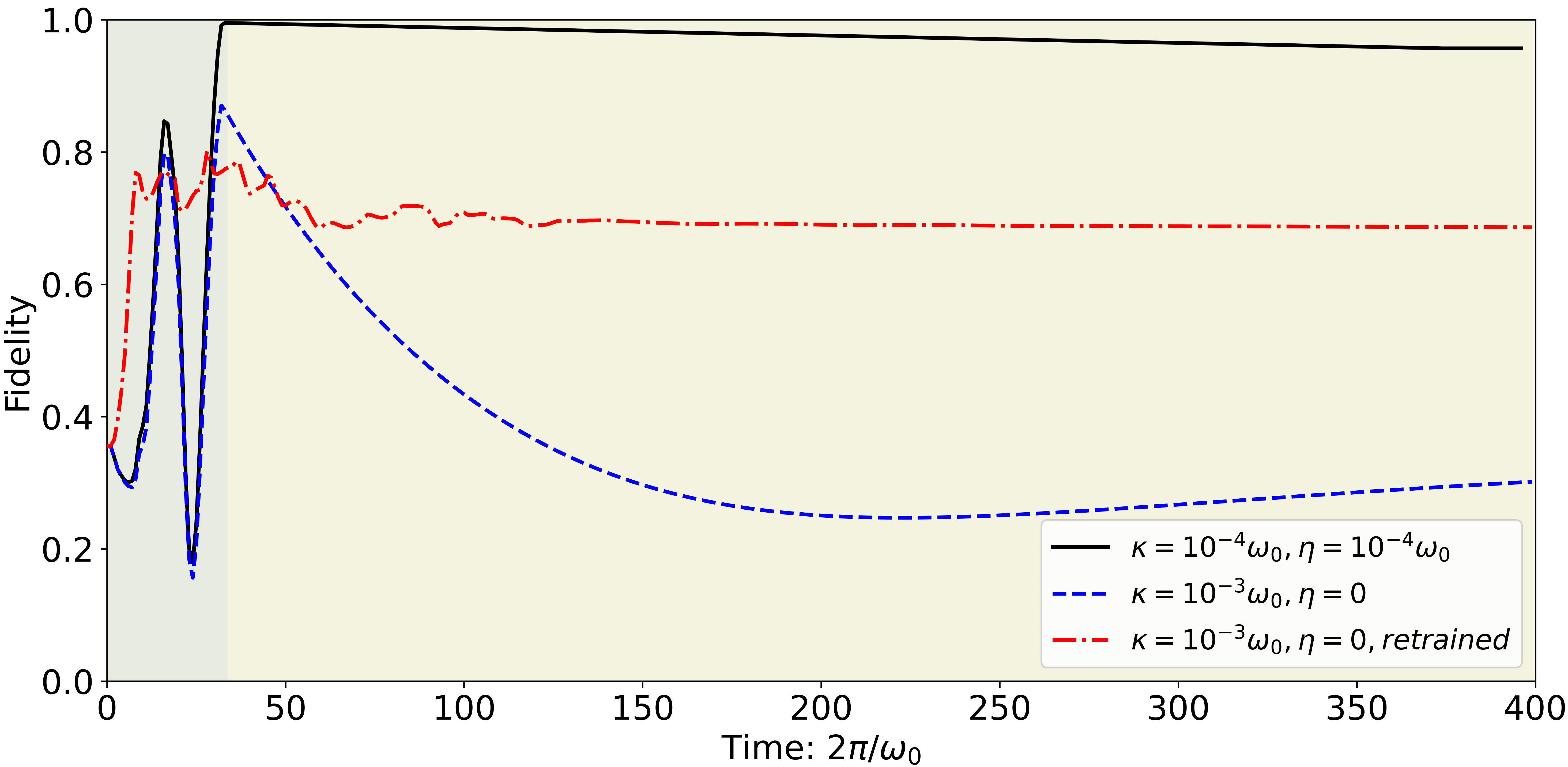}

         \caption{Stroboscopic 
time evolution of fidelity of the prepared state under various environment noise levels with our reinforcement learning method. The black solid line shows the 
 fidelity evolution at a relatively high noise rate with our original trained policy, i.e.  $\kappa=10^{-4}\omega_0, \eta=10^{-4}\omega_0$, where two stages are defined with various background colors. The blue dashed line shows the fidelity evolution  at a extremely  high noise rate with our original trained policy,  i.e. $\kappa=10^{-3}\omega_0, \eta=0$,  where the fidelity of the generated state in stage II  drops vastly.     The red dashed-dotted line shows the fidelity evolution with our retrained model at the same  noise level of the blue dashed line case.}
         \label{fig:noise-study}
\end{figure}

\begin{figure}
    \centering
\includegraphics[width=1.02\columnwidth]{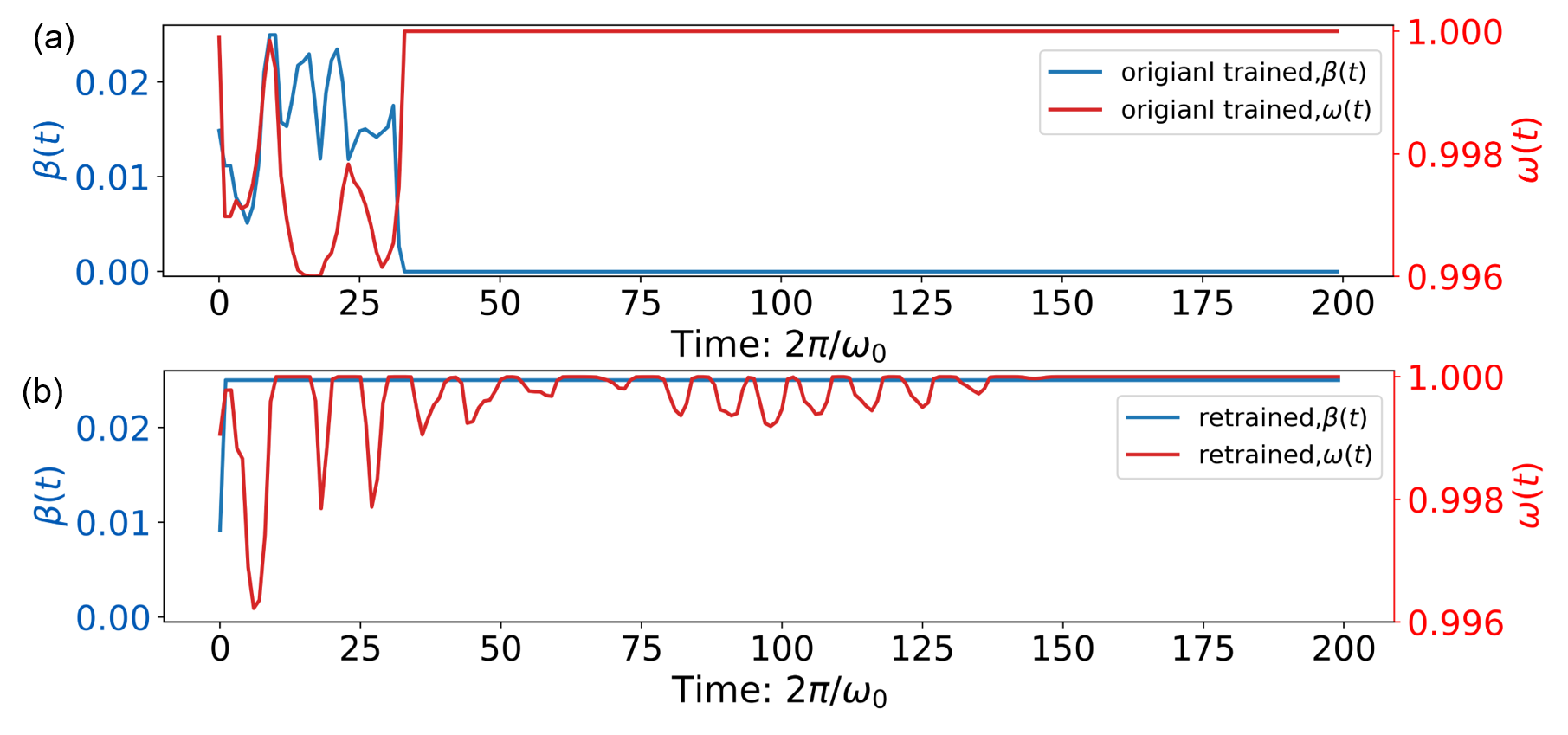}
         \caption{The reinforcement learned control sequences of $\beta(t)$ and $\omega(t)$ obtained with our (a) original  trained network and (b)  the  retrained network, corresponding to  the blue dashed line and the red dashed-dotted line  in Fig.~\ref{fig:noise-study} respectively.}
         \label{fig:beta-omega-red-curve-inFig4}
\end{figure}

\section{Results}
Here, we present an efficient  robust bosonic codes preparation scheme empowered by reinforcement learning and Floquet engineering.  The Floquet engineering has been proposed for the preparation of bosonic codes under the adiabatic ramp process, where the temporal variation of the amplitude $\beta(t)$ and $\omega(t)$ follows  a heuristic Sigmod  formula~\cite{guo-PhysRevLett.132.023602,Arne-PhysRevLett.132.130605}. 
Despite the theoretical feasibility of the Sigmod formula, it requires quite a long evolution time (usually several thousands of driving periods) for the preparation of the target state with a satisfied fidelity.  For more than half of the evolution time, the system oscillates regularly without increasing fidelity~\cite{guo-PhysRevLett.132.023602}, as shown by the inset of Fig.~\ref{fig:evolution-process}~(b). Except the long preparation time, the adiabatic ramp process is also vulnerable to the environment noise. Therefore, a fast and noise-resilience state preparation  method is in high demand for the practical application of quantum bosonic codes.

Our proposed solution introduces the reinforcement learning during the Floquet engineering process using the popular TD3 policy gradient algorithm~\cite{fujimoto2018addressing}. 
During our TD3 policy gradient algorithm, the agent learns to optimize the value of $\beta(t)$ and $\omega(t)$ to ensure the system evolves to the target state with high fidelity within a much shorter time period (about $1\%$ of the originally required time). As shown in Fig.~\ref{fig:evolution-process}~(b), compared to the total preparation time $5000\times2\pi/\omega_0$ required by the adiabatic ramp in Ref.~\cite{guo-PhysRevLett.132.023602}, cf. inset of Fig.~\ref{fig:evolution-process}~(b), our method manages to prepare the  target 4-fold rotational target bosonic state  in around $50\times2\pi/\omega_0$  with the same level of high fidelity ($\mathcal{F}\sim 0.998$).  We show the machine learned driving field parameters $\omega (t)$ and $\beta(t)$ in the whole process of generating 4-fold rotational target bosonic state in Fig.~\ref{fig:evolution-process}~(c). Specifically, the sequence of adjustment actions that the agent finds behaves in the following way: the $\beta(t)$ is large initially and oscillates frequently in the early stage, and then converges to zero as the prepared state fidelity converges to 1  as shown in~Fig.~\ref{fig:evolution-process}~(c). In the meantime, the $\omega(t)$ oscillates synchronously and converges to a fixed number ($1$ in our case), which could be other constants as the $\beta$ is zero in these circumstances (therefore the value of $\omega$ does not affect the evolution) as shown in~Fig.~\ref{fig:evolution-process}~(c).  In Fig.~\ref{fig:evolution-process}~(a), we plot several snapshots of Husmi Q-functions of prepared states to show 
the dynamical formation process of our 4-fold rotational bosonic state.

We further investigate the robustness of our method by considering the photon loss and the photon dephasing in the experimentally relevant scenario occurring as shown in~Fig.~\ref{fig:noise-test}. In the experiment of superconducting circuits, the photon loss rate $\kappa$ and the dephasing rates $\eta$ are both much smaller than the typical frequency of the oscillations in the rotating frame (set by driving strength $\beta$). In our simulations, we set the maximum value of  $\beta/\omega_0=0.025$ to satisfy the rotating wave approximation (RWA) and investigate the $4$-fold cat state preparation with various values of $\kappa$ and $\eta$ as shown in Fig.~\ref{fig:noise-test}, where the results in all levels noisy environments are obtained using the network trained with the lowest noise model, i.e. the case with ${\kappa = 10^{-7}\omega_0}$, 
${\eta=10^{-7}\omega_0}$.  Our simulations demonstrate the noise robustness and practicality of our model, while the original heuristic method is noise sensitive, and the prepared state fidelity drops vastly as the noise level increases~\cite{guo-PhysRevLett.132.023602}. Specifically, the neural network for the state preparation process trained in the low-noise  environment with  $\eta=10^{-7}\omega_0, \kappa=10^{-7}\omega_0$  can prepare a high-quality state even in the relatively high-noise environment  without retraining, such as the  high-noise environments with  $\eta=10^{-4}\omega_0, \kappa=10^{-4}\omega_0$. 

Furthermore, in the environment with the extremely strong noise where the prepared state has quite low fidelity (around $0.2$) with the model trained in low noise environment, our results can be further improved by retraining the neural network, as shown by the red dash dotted line in Fig.~\ref{fig:noise-study}.  
With a neural network trained under the low-noise environment with ${\kappa = 10^{-7}\omega_0}$, ${\eta=10^{-7}\omega_0}$, the fidelity during the 4-fold rotational state generation process in the relatively high noise environment with ${\kappa = 10^{-4}\omega_0}$, ${\eta=10^{-4}\omega_0}$ is shown by the black solid line in  Fig.~\ref{fig:noise-study}. In this case, the fidelity of the prepared state reaches the highest value ($\sim 0.998$) at around $t=50\times 2\pi/\omega_0$ (stage I), then drops with a low constant speed(stage II). 
For the extremely high noise environments, the fidelity of the prepared state using our low-noise environment trained network drops vastly in stage II,  such as a fidelity around $0.2$ for a noise environment with $\kappa=10^{-3}\omega_0$ as shown by the blue dashed line in Fig.~\ref{fig:noise-study}. 
Within such high noise environments, we can retrain our neural network to achieve a high fidelity.  Our neural network manages to learn a noise resistance policy, to stabilize the fidelity in stage II, as shown by the red dash-dotted line in Fig.~\ref{fig:noise-study}.

The learned control sequences, i.e. $\beta(t)$ and $\omega(t)$, for our extremely high noise environment with ${\kappa = 10^{-3}\omega_0}$, ${\eta=0}$ with original  neural network trained in environment with ${\kappa = 10^{-7}\omega_0}$, ${\eta=10^{-7}\omega_0}$  is shown  in~Fig.~\ref{fig:beta-omega-red-curve-inFig4}(a). The learned control sequences  for our extremely noise environment of ${\kappa = 10^{-3}\omega_0}$, ${\eta=0}$ via retraining our neural network is shown in~Fig.~\ref{fig:beta-omega-red-curve-inFig4} (b). Within our retrained learned policy, both the amplitude and frequency of the driving filed converges to the high values, compared with the behavior of $\beta$ converging to zero driving field in the original learned policy. This phenomena could be interpreted as  that the driving field constantly contributes to the noise resulted error  corrections.

\begin{figure*}
    \centering
\includegraphics[width=1.4\columnwidth]{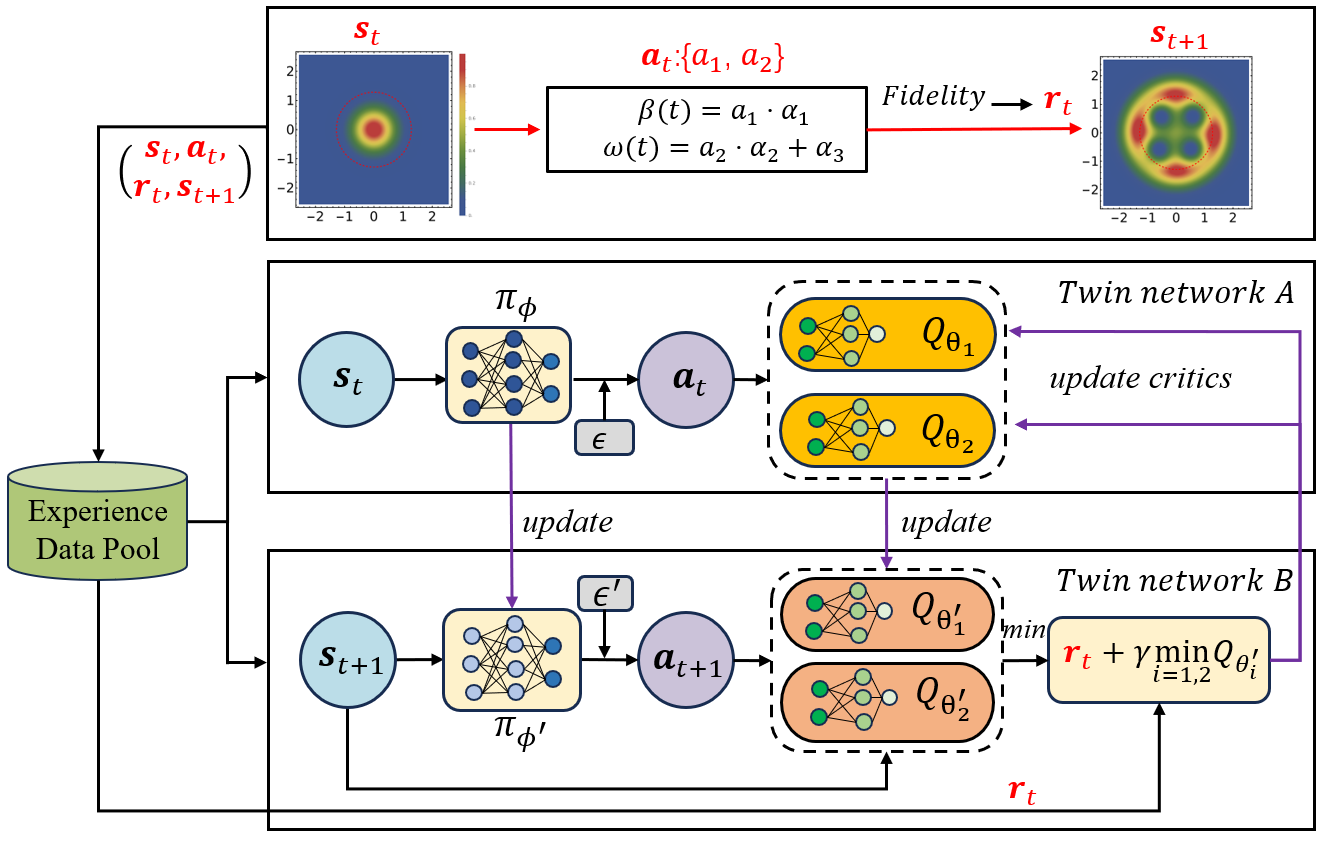}
         \caption{Reinforcement learning algorithm flowchart for the bosonic code state preparation. The Red arrows represents prediction process, the black and purple arrows represent  the training process, where the purple arrows make policy gradient updates more stable and efficient. Here, $s_t$ is current state, $a_t$ is current action generated by policy network $\pi_\phi$, $r_t$ is instantaneous reward,  $\epsilon$ is Gaussian noise, $\pi_\phi$ is policy network, $Q_\theta$ is action value.}
         \label{fig:Algorithm Flowchart}
\end{figure*}

\begin{algorithm}[t]
\caption{TD3 for preparation of rotation-symmetry codes}
\label{alg:td3}
\SetAlgoLined
\DontPrintSemicolon
Initialize critic networks \( Q_{\theta_1}, Q_{\theta_2} \), and actor network \( \pi_{\phi} \) with random parameters \( \theta_1, \theta_2, \phi \)\;
Initialize target networks \( \theta'_1 \leftarrow \theta_1 \), \( \theta'_2 \leftarrow \theta_2 \), \( \phi' \leftarrow \phi \)\;
Initialize experience data pool \( D \)\;
Initialize exploration noise variance \( \sigma \) and noise attenuation factor constant \(\delta\)\;
Initialize \(a_t:\)\{\(a_1,a_2\) \}, the amplitude $\beta$ and phase $\omega$, \(\alpha_1\)=0.025,\(\alpha_2\)=0.004,\(\alpha_3\)=0.996\;
Initialize target policy random noise \(\tilde{\sigma}\), noise clipping threshold constant \(c\) \;
Initialize critic networks and actor network update proportion constant \(\tau\)\;
\For{\( t = 1 \) \KwTo \( T \)}{
    \eIf{$ t < T_0 $}{
        Select action ${a_t:\{{a_1,a_2}}\}$  with exploration:\;

      ${a} \sim \mathcal{N}(0,1)$\;
 
    }{

        Select action ${a_t:\{{a_1,a_2}}\}$ with exploration noise \( \epsilon \): \; 
         \( a_t \sim \pi_{\phi}(s) + \epsilon \), \( \epsilon \sim \mathcal{N}(0, \sigma) \)\;
          \( \sigma = max(\sigma-\delta\) , 0)\;
    }
   \(\beta(t)=a_1 \cdot \alpha_1 \)\;
    \(\omega(t)=a_2 \cdot \alpha_2 + \alpha_3   \)\;
     Observe new state \( s_{t+1} \) and Calculate fidelity\;
        Using fidelity to calculate reward \( r_t \) by reward function \(\mathcal{F}(t,Fidelity)\)\;
        Store transition tuple \( (s_t, a_t, r_t, s_{t+1}) \) in \( D \)\;
    Sample mini-batch of \( N \) transitions \( (s_t, a_t, r_t, s_{t+1}) \) from \( D \)\;
    \( \tilde{a_{t+1}} \leftarrow \pi_{\phi'}(s_{t+1}) + \epsilon' \), \( \epsilon' \sim \mathrm{clip}(\mathcal{N}(0, \tilde{\sigma}), -c, c) \)\;
    \( y \leftarrow r + \gamma \min\limits_{i=1,2} Q_{\theta'_i}(s_{t+1}, \tilde{a_{t+1}}) \)\;
    Update critics: \( \theta_i \leftarrow \min_{\theta_i} N^{-1} \sum (y - Q_{\theta_i}(s_t,a_t))^{2} \)\;
    
    \If{$ t \mod d = 0 $}{
        Update \( \phi \) by deterministic policy gradient:\;
        \( \nabla_{\phi}J(\phi) = N^{-1} \sum \nabla_{a} Q_{\theta_1}(s_t,a)\big|_{a=\pi_{\phi}(s_t)} \nabla_{\phi} \pi_{\phi}(s_t) \)\;
        Update target networks:\;
        \( \theta'_i \leftarrow \tau\theta_i + (1-\tau)\theta'_i \)\;
        \( \phi' \leftarrow \tau\phi + (1-\tau)\phi' \)\;
    }
}
\end{algorithm}

\section{Methods} 
In our reinforcement learning scheme with noises, the environment represents the quantum dissipative dynamics. The time evolution of the density matrix in the presence of weak photon loss and pure dephasing is described by the Lindblad master equation~\cite{osti_4208943},
\begin{align}
 \frac{d}{dt}\rho(t)=&-\frac{\text{i}}{\lambda}[\hat{\mathcal{H}}(t),\rho(t)] +\kappa \left(\hat{a}\rho(t)\hat{a}^\dagger-\frac{1}{2}\{\hat{a}^\dagger\hat{a},\rho (t)\}\right)\nonumber\\ &+\eta\left(\hat{a}^\dagger\hat{a}\rho(t)\hat{a}^\dagger\hat{a}-\frac{1}{2}\{(\hat{a}^\dagger\hat{a})^2,\rho(t)\}\right).
 \label{eq:lindbald-eq}
\end{align}
Here, $\{\hat{A},\hat{B}\}\equiv\hat{A}\hat{B}+\hat{B}\hat{A}$ is the anticommutator, $\kappa$ is the single-photon loss rate and $\eta$ is the dephasing rate. The $\hat{\mathcal{H}}(t)$ is the Hamiltonian defined in Eq.~(\ref{eq:RL-hamiltonian}), with the $\beta(t)$ and $\omega(t)$ being controlled via the TD3 reinforcement learning method. Our Lindbald master equation in  Eq.~(\ref{eq:lindbald-eq}) is valid for any weakly nonlinear cavity~\cite{osti_4208943} with high quality factor, i.e., $\beta\ll\omega_0$ and $\kappa,\eta\ll\omega_0$. In a realistic parameter regime $\eta|\alpha_e|^2\ll\kappa\ll 2 q^2\beta\ll\omega_0$ with $\alpha_e=(x_e+ip_e)/\sqrt{2\lambda}$ as  the minima of the target Hamiltonian given by Eq.~(\ref{eq:target-hamiltonian}), the oscillator tends to relax into the groundstate manifold of the bosonic code Floquet Hamiltonian.

During our reinforcement learning process, the updated control for the quantum systems is decided by an agent represented by a twin neural network as shown in Fig.~\ref{fig:Algorithm Flowchart}. 
The update logic is referred to as a policy in our reinforcement learning and is determined based on the quality of the current system state.  The quality of the system state is evaluated  by the fidelity between  the current prepared state $\rho(t)$ and the target state $\rho_0$, defined as 
\begin{equation}
    \mathcal{F}[\rho_0,\rho(t)]=\text{Tr}\sqrt{\rho_0^{1/2}\rho(t)\rho_0^{1/2}}=\sqrt{\text{Tr}[\rho(t)\rho_0]}.
    \label{eq:fidelity}
\end{equation}
Specifically, the policy in our TD3 reinforcement learning algorithm consists of two processes, i.e.  training (shown in  Alg.~\ref{alg:td3}) and prediction as shown in  Figs.~(\ref{fig:sketch},\ref{fig:Algorithm Flowchart}). During the beginning of the training process ($t<T_0$ in Alg.~\ref{alg:td3}),  an  experience pool is generated by randomly selected actions. 
Afterwards, the policy network $\pi_\phi$ and the action-value network are trained to update simultaneously. 

For the update of policy network $\pi_\phi$,  using a randomly selected batch size data from the experience pool, 
the policy ${\phi}$ is updated through a  deterministic policy gradient descent algorithm $\nabla_{\phi} J(\phi)$, which aims to maximize the   target function   $J(\phi)$ defined via
\begin{equation}
      J(\phi) = \mathbb{E}_{s \sim \mathcal{D}} [ Q_\theta(s, a) ],\ a=\pi_\phi(s)
\end{equation} 
with $\mathcal{D}$ the data in experience data pool and the expected action value given by
\begin{equation}
Q_{\theta}(s_t,a_t) = \mathbb{E} \left[ r_t + \gamma \, \min_{i=1,2} Q_{\theta_i'}(s_{t+1}, \pi_{\phi'}(s_{t+1}) + \epsilon') \right].
\label{eq:Q-theta}
\end{equation}
Here, $r_t$ is instantaneous reward, $s_t$ is current state, 
$a=\pi_\phi(s)$ is the action generated by  policy network $\pi_\phi$, $\gamma$ is discounted factor, $\epsilon'$ is Gaussian noise. 
Therefore, the formula of gradient descent algorithms presents,
\begin{equation}
\nabla_{\phi} J(\phi) = \mathbb{E}_{s \sim {D}} \left[ \nabla_a Q_\theta(s, a=\pi_\phi(s)) \nabla_{\phi} \pi_{\phi}(s) \right].
\end{equation}

 Optimization of the expected action value in Eq.~(\ref{eq:Q-theta}) is equivalent to  maximizing the instantaneous reward $r_t$, which is designed to accelerate fidelity convergence while ensuring steady-state stability. The reward $r_t$ is defined based on the fidelity $\mathcal{F}$ with the following formula,
\begin{equation}
r_t= 10\mathcal{F}_t^5
- 0.01\sqrt{t}
+ R_{\text{success}}
+ R_{\text{stab}}
\label{eq:reward}
\end{equation}
with
\begin{align}
R_{\text{success}} &= \begin{cases}
    \dfrac{0.1}{T_0}(5000-t) \sum_{k=0}^{T_0-1} \mathcal{F}_{t-k} & c_t\geq 700 \nonumber\\
    0 & \text{otherwise}
\end{cases} \\
R_{\text{stab}} &= \begin{cases} 
    0.5c_t & \mathcal{F}_t \geq 0.85 \wedge c_t < 20 \\
    10\log(c_t) & \mathcal{F}_t \geq 0.85 \wedge c_t >= 20 \\
    0 & \mathcal{F}_t < 0.85 \wedge c_t = 0 \\
    -0.2c_t & \mathcal{F}_t < 0.85 \wedge c_t > 0
\end{cases}\nonumber \\
c_t &= \begin{cases}
    c_{t-1} + 1 & \mathcal{F}_{t-1} \geq 0.85 \\
    0 & \text{otherwise},
\end{cases}\nonumber
\end{align}
The first term on the right-hand side of Eq.~(\ref{eq:reward}) encourages a high-fidelity state, the second term represents the time penalty which  discourages prolonged evolution. 
The third term offers the high reward to the actions achieving high fidelity within short time, and  maintaining a high value of the average fidelity during the end stage of the evolution. The fourth term represents the stability reward, which promotes the policy to converge as soon as possible and enables the fidelity to stabilize. 

During the training process, except the aforementioned policy network training, i.e. the update of policy network $\pi_\phi$ and action-value network $Q_\theta$ is simultaneously implemented  by minimizing temporal difference $L(\theta)$ given by
\begin{equation}
    \begin{aligned}
        L(\theta) &= \mathbb{E}_{(s_t,a_t,r_t,s_{t+1}) \sim \mathcal{D}} \Bigl[Q_{\theta}(s_t,a_t) \\
            & - \left( r_t + \gamma Q_{\theta'}\bigl(s_{t+1}, \pi_{\phi'}(s_{t+1})\bigr) \right) \Bigr]^2,
    \end{aligned}
\end{equation}
where $\theta'$ and $\phi'$ are parameters in  action-value  network and policy network respectively. During the prediction process, the action of the agent   is the driving field with the parameters  $\beta(t)$ and ${\omega(t)}$ extracted from the policy network  ${\pi_{\phi}}$,  and we manage to prepare the high-fidelity 4-fold rotation bosonic code state efficiently.

Our experiments are implemented on a computer configured with two AMD EPYC 9554 CPUs, four NVIDIA 4090 GPUs, 512GB of memory, a Ubuntu 20.04 system, Python 3.8.13, the torchquantum 0.1.5 framework. 
Each experiment was implemented five times, with the seed values set to [0,1,2,3,4] for each of the experiments.

\section{Conclusion}
In summary, the preparation of bosonic codes remains one of the central challenges in the study of continuous-variable quantum computing, primarily due to the stringent requirements on coherence time and control precision. Floquet engineering combined with adiabatic ramp protocols developed in recent years~\cite{Arne-PhysRevLett.132.130605,guo-PhysRevLett.132.023602}, while conceptually straightforward, often demand exceedingly long evolution times and are highly susceptible to various noise sources, such as photon loss and dephasing.

In this work, we have introduced a reinforcement machine learning–assisted Floquet engineering scheme that substantially enhances both the efficiency and robustness of bosonic code preparation. Remarkably, our method reduces the required preparation time to approximately one percent of that in the previously proposed adiabatic approaches, while maintaining high-fidelity state generation even under strong noise conditions. This demonstrates not only the effectiveness of Floquet-based control for continuous-variable systems, but also the capability of machine learning techniques to discover nontrivial optimal control strategies that surpass conventional human-designed protocols.

Our results highlight the potential of combining artificial intelligence and quantum control as a powerful paradigm for realizing fault-tolerant bosonic quantum computation. We anticipate that this framework can be further extended to a broad class of quantum platforms and noise models, providing a scalable route toward practical implementations of quantum error-correcting codes in near-term quantum hardware.

\section{Data availability}
The data that support the findings of this study are available from the
corresponding author upon request.

\section{Code availability}
The custom codes for this study that support the findings are available
from the corresponding authors upon request.

\section{Author contributions}
Wei-Wei Zhang developed the theoretical framework. Lingzhen Guo contributed to the Floquet engineering of bosonic codes.   Zheping Wu and Wei-Wei Zhang conceived the experimental scheme. Zheping Wu conducted the code realization of this work. All authors performed the data analysis and discussed the results, and contributed to the preparation of the manuscript.

\section{Competing interests}
The authors declare no competing interests.

\section{Additional information}
Correspondence and requests for materials should be addressed to
Wei-Wei Zhang.

\section{Acknowledgments}
We acknowledge support from the National Natural Science Foundation of China~(Grant  No.~62571434, 12475025).

\bibliography{ref-rl-gkp}

\end{document}